\documentclass[12pt,a4paper]{article}
\usepackage{epsfig}
\usepackage{graphicx}
\title{Study of the single neutral top-pion production process at $\gamma\gamma$ collider}
\author{Xuelei Wang$^{a,b}$, Xiaoxue Wang$^{b,c}$\\
{\small a: CCAST(World Laboratory) P.O. BOX 8730. B.J.
100080, P.R.China}\\
 {\small b: College of Physics and Information
Engineering,}\\
\small{Henan Normal University, Xinxiang  453007, P.R.China}\\
{\small c: College of Physics and Chemistry,}\\
\small{Henan Polytechnic University, Jiaozuo, Henan, 454000,
P.R.China}
\thanks{E-mail:wangxuelei@sina.com}
\thanks{Mailing address} }
\begin{document}
\maketitle

\begin{abstract}
\hspace{5mm} $\gamma\gamma\rightarrow \Pi_t^0$ is the major
production mechanism of neutral top-pion at the linear colliders.
In this paper, we calculate the cross section of the process
$\gamma\gamma \rightarrow \Pi^0_t$ and discuss the potential to
observe the neutral top-pion via its various decay modes at the
planned ILC. The study show that, among the various neutral
top-pion production processes at the linear colliders, the cross
section of $\gamma\gamma\rightarrow \Pi_t^0$ is the largest one
which can reach the level of $10^1-10^2$ fb. Due to the existence
of the tree-level flavor-changing coupling $\Pi_t^0 t\bar{c}$,
$\gamma\gamma\rightarrow \Pi_t^0 \rightarrow t\bar{c}$ can provide
enough number of typical signals to identify the neutral top-pion
with the clean SM background. Therefore, the process
$\gamma\gamma\rightarrow \Pi_t^0$ play an important role in
searching for the neutral top-pion and test the TC2 model.
\end {abstract}
PACS number(s): 12.60Nz,14.80.Mz,12.15.Lk,14.65.Ha
\newpage

\section{Introduction}
 The mechanism that governs the electroweak
symmetry breaking(EWSB) is at present the largest unknown in the
Standard Model(SM). In order to solve the problem of EWSB, some
new physics models beyond the SM have been proposed, such as:
supersymmetry (SUSY) and dynamical EWSB mechanism concerning new
strong interactions, etc. An important goal of studies at the next
generation of positron-electron($e^+e^-$) colliders is the proper
understanding of the EWSB, and the discovery of new physics beyond
the SM. If the new particles or interactions will be directly
discovered at the Tevatron and LHC, the linear colliders will then
play a crucial role in the detailed and thorough study of these
new phenomena and in the reconstruction of the underlying
fundamental theories. The International Linear
Collider(ILC)\cite{ILC} is a worldwide consensus linear
accelerator $e^+e^-$ collisions in the energy range of 500 GeV to
above TeV. By Compton backscattering of laser photons off the
electron and positron beams, one can produce high luminosity
$\gamma\gamma$ collisions with a wide spectrum of $\gamma\gamma$
center-of-mass energy. In addition, a high degree of circular
polarization for each of the colliding photons can be achieved by
polarizing the incoming electron and positron beams and the laser
beams\cite{polarization}. Photon colliders have distinct
advantages in searches for and measurements of new physics
objects. In general, phenomena in $e^+e^-$ and
$\gamma\gamma,\gamma e$ collisions are similar because the same
particles can be produced. However, the reactions are different
and the photon colliders often give complementary information.
Some phenomena can be studied better at photon colliders due to
higher statistical accuracy. On the other hand, the cross sections
of some processes in $\gamma\gamma$ collisions are larger than
those in $e^+e^-$ collisions. In the hunt for physics beyond the
SM, only small signs may be visible, therefore, the photon
colliders provide optimal conditions for searching for the new
physics.

Among the new physics models, technicolor(TC) model is a promising
candidate of dynamical theories\cite{TC}. However, it is hard for
technicolor to generate the fermion masses, specially, the heavy
top quark mass. In order to overcome the shortcoming of the simple
TC model and explain the large mass problem of the top quark, an
interesting TC model, called the topcolor assisted technicolor
(TC2) model, is proposed \cite{Hill, Lane,Cvetic,Buchalla} which
gives a reasonable explanation of the EWSB and heavy top quark
mass. In the TC2 model, the topcolor interaction makes small
contribution to the EWSB, and gives rise to the main part of the
top quark mass $(1-\varepsilon)m_{t}$ with a model dependant
parameter $0.03\leq\varepsilon\leq 0.1$\cite{Buchalla}. The
technicolor interaction plays a
 main role in the breaking of electroweak gauge symmetry. To
 account for the explicit breaking of quark and lepton flavor
 symmetries, the extended technicolor(ETC) was invented.
  The ETC interaction gives rise to
 the masses of the ordinary fermions including a very small portion
 of the top quark mass $\varepsilon m_{t}$. This kind of model predicts
 three CP odd top-pions $(\Pi^{0}_{t}$,$\Pi^{\pm}_{t})$ with large
 Yukawa couplings to the
 third family. The LEP-SLD precision measurement data of $R_b$ give a severe
 constraint on the mass of the charged top-pion\cite{constraint}.
 Even in an optimistic estimate the mass of the charged top-pion should
  be larger than 220 GeV. Because the top-pions are the typical physical
 particles of the TC2 model, the observation of top-pions can be regarded as
 the direct evidence of the TC2 model. The study of the various top-pion
production mechanisms is well motivated which can offer the useful
instruction to search for the top-pions. Via the $e^+e^-$
collisions, the neutral top-pion can be produced via the processes
$e^+e^-\rightarrow f\bar{f}\Pi_t^0(f=u,d,c,s,t,b,e,\mu,\tau),
e^+e^-\rightarrow t\bar{c}\Pi^0_t,e^+e^-\rightarrow
\Pi_t^0Z,\Pi_t^0\gamma$. The main charged top-pion production
mechanism are $e^+e^-\rightarrow t\bar{b}\Pi_t^-$ and
$e^+e^-\rightarrow W^+\Pi_t^-$, these processes have been
systematically studied\cite{wang1,yue,du}. The photon colliders
will offer another chance to search for the top-pions. We have
studied the potential to discover the neutral top-pion via the
processes $\gamma\gamma \rightarrow t\bar{t}\Pi_t^0, \gamma\gamma
\rightarrow t\bar{c}\Pi_t^0, e^-\gamma \rightarrow
e^-\Pi_t^0$\cite{wang2,wang3}. In general, the cross sections of
above processes are at the level of $10^0-10^1$ fb, and it is
promising to observe the top-pions at future linear colliders with
high luminosity. Also, the charged top-pion production processes
at the photon collider have been studies in the reference
\cite{He}. On the other hand, the neutral top-pion can also be
produced as s-channel resonances in the process $\gamma\gamma
\rightarrow \Pi^0_t$ through top quark triangle loop and thus the
full photon beam energy can be used to produce the heavy particle.
The similar work has been done in the framework of a rescaled QCD
models and the loe-scale TC model\cite{similar}. In this paper, we
calculate the cross section of $\gamma\gamma \rightarrow \Pi^0_t$
and study the possibility to observe the neutral top-pion via its
various decay modes. We find that, with large cross section,
$\gamma\gamma \rightarrow \Pi^0_t$ can provide a unique way to
identify the neutral top-pion with high precision. Among the
various decay modes of $\Pi_t^0$, $\Pi_t^0\rightarrow t\bar{c}$ is
a typical one due to the existence of the tree-level coupling
$t\bar{c}\Pi_t^0$. Because there exists GIM(Glashow, Iliopoulos,
and Maiani) mechanism in the SM, $t\bar{c}$ production rate in the
SM is too small to observe $t\bar{c}$, the observable signal of
$t\bar{c}$ might be a sound evidence of the new physics. So,
$\Pi_t^0\rightarrow t\bar{c}$ is the ideal mode to observe
$\Pi_t^0$. $t\bar{c}$ production processes via photon-photon
collision have been studied in some new physics models(The Two
Higgs Doublet Model(Model III), Minimal Supersymmetic Standard
Model(MSSM) and TC2 model\cite{rrtc}. The study shows that the
production rate of $t\bar{c}$ can reach the observable level in
these new physics models. In our paper, the possibility to observe
the neutral top-pion via other modes is also systematically
studied.

 The outline of the paper is as
follows: In the next section, we describe the details of our
calculation and give some analytic formulae. In the third section,
we discuss the numerical results and give some conclusions.

\section{The cross section of  $\gamma\gamma \rightarrow \Pi^0_t$}
As it is well known, the couplings of the top-pions to the three
 family fermions are non-universal. The top-pions have large Yukawa
 couplings to the third family and can induce large flavor
 changing couplings. The couplings of the neutral top-pion
 to quarks can be written
 as\cite{He}:
\begin{eqnarray}
i\frac{m_ttan\beta}{v_w}[K_{UR}^{tt}{K_{UL}^{tt}}^{\ast}
\bar{t_L}t_R\Pi_t^0+K_{UR}^{tc}{K_{UL}^{tt}}^{\ast}
\bar{t_L}c_R\Pi_t^0+\frac{m_b^*}{m_t}K^{bb}_{DL}\bar{b}_Lb_R\Pi_t^0+h.c.].
\end{eqnarray}
 Where, $tan\beta=\sqrt{v_w^2/v_t^2-1}$, $v_w=246$ GeV is the electroweak symmetry breaking
 scale, and $v_t=60 \sim 100$ GeV is the top-pion decay constant.
 $m^{*}_{b}(m^{*}_{b}\approx \frac{3km_{t}}{8\pi^{2}}
 \sim 6.6k$ GeV) is the part of b-quark mass induced by the instanton
 and k ranges from 1 to $10^{-1}$ as in
QCD, in our calculation, we take k=0.5 as a typical example. The
factor
 $tan\beta$ reflects the effect
 of the mixing between the top-pions and the would-be goldstone
 bosons. $K^{tt}_{UL}$, $K^{bb}_{DL}$, $K^{tt}_{UR}$ and $K^{tc}_{UR}$ are the elements
of the rotation matrices $K_{L,R}$ which are needed for
diagonalizing the quark mass matrices. The matrix elements are
given as
 \begin{eqnarray}
K^{tt}_{UL}\approx K^{bb}_{DL}\approx 1, \hspace{1.5cm} K^{tt}_{UR}=1-\varepsilon,\\
K_{UR}^{tc}=\sqrt{1-{K_{UR}^{tt}}^2-{K_{UR}^{tu}}^2}\leq\sqrt{2\varepsilon-\varepsilon^2}.
\end{eqnarray}
In this paper, we take
$K_{UR}^{tc}=\sqrt{2\varepsilon-\varepsilon^2}$ and typically take
$\varepsilon=0.03,0.06,0.1$.

With the coupling $t\bar{t}\Pi_t^0$, $\Pi_t^0$ couples to a pair
of photon via top quark triangle loop(the contribution of bottom
quark loop is much smaller than that of top quark loop, we ignore
such contribution here). We explicitly calculate the top quark
triangle loop and obtain the effective coupling of
$\Pi_t^0-\gamma-\gamma$ as:
\begin{equation}
\frac{8im_t^2\tan\beta(1-\varepsilon)\alpha_e}{9v_w\pi}\varepsilon_{\mu\nu\rho\sigma}
p_3^{\rho}p_2^{\sigma}C_0,
\end{equation}
where $C_0=(-p_2,p_3,m_t,m_t,m_t)$ is the standard three-point
Feynman integral with $p_3$ donating the momentum of $\Pi_t^0$ and
$p_2$ denoting the momentum of an incoming photon. Single
$\Pi_t^0$ can be produced via photon-photon fusion process
$\gamma\gamma\rightarrow\Pi_{t}^{0}$.

The lowest order parton cross section can be expressed by the
decay width of $\Pi_t^0\rightarrow\gamma\gamma$\cite{cross}:
\begin{equation}
\hat{\sigma}_{\gamma\gamma\rightarrow\Pi_t^0}(\hat{s})=\sigma^0
M_{\Pi}^2\delta(\hat{s}-M_{\Pi}^2),
\end{equation}
\begin{equation}
\sigma^0=\frac{\pi^2}{8M_{\Pi}^3}\Gamma(\Pi_t^0\rightarrow\gamma\gamma),
\end{equation}
\begin{equation}
\Gamma(\Pi_t^0\rightarrow\gamma\gamma)=\frac{2m_t^4\tan^2\beta(1-\varepsilon)^2
\alpha_e^2M_{\Pi}^3}{81\pi^3 v_{w}^2}|C_0|^2,
\end{equation}
where $\hat{s}$ is the $\gamma\gamma$ center-of-mass(c.m.) energy
squared and $M_{\Pi}$ denotes the mass of the neutral top-pion.
The $\delta$ distribution can be approximately substituted by the
Breit-Wigner form for zero-width $\delta$
distribution\cite{cross}:
\begin{equation}
\delta(\hat{s}-M_{\Pi}^2)\rightarrow\frac{1}{\pi}\cdot\frac{\hat{s}\Gamma_{\Pi}/M_{\Pi}}
{(\hat{s}-M_{\Pi}^2)^2 +(\hat{s}\Gamma_{\Pi}/M_{\Pi})^2},
\end{equation}
 and
changing kinematical factors $M_{\Pi}^2\rightarrow\hat{s}$
appropriately. $\Gamma_{\Pi}$ is the total decay width of
$\Pi_t^0$ which can be obtained by summing the decay widths of all
the
 decay modes. The possible decay modes of $\Pi_t^0$ are
$\Pi_t^0\rightarrow\:{t\bar{t}},\:{t\bar{c}},{b\bar{b}},\:\gamma\gamma,\:{gg},
\:\gamma{Z}$(If $M_{\Pi}<2m_t$, $\Pi_t^0\rightarrow t\bar{t}$ is
forbidden). Their decay widths can be easily obtained as:

$$\Gamma_{\Pi_t^0\rightarrow{t\bar{t}}}=\frac{3(1-\varepsilon)^2m_t^2\tan^2\beta
 M_{\Pi}}{8\pi v_{w}^2}\sqrt{1-\frac{4 m_{t}^2}
{M_{\Pi}^2}},$$

$$\Gamma_{\Pi_t^0\rightarrow{t\bar{c}}}=\frac{3(2\varepsilon-\varepsilon^{2})m_t^2\tan^2\beta
{(M_{\Pi}^2-m_{t}^2)}^2 }{8\pi v_{w}^2 {M_{\Pi}^3}},$$

$$\Gamma_{\Pi_t^0\rightarrow{b\bar{b}}}=\frac{3{m_b^\ast}^2\tan^2\beta
M_{\Pi}}{8\pi v_{w}^2}\sqrt{1-\frac{4 m_{b}^2} {M_{\Pi}^2}},$$

$$\Gamma_{\Pi_t^0\rightarrow{gg}}=\frac{(1-\varepsilon)^2m_t^4M_{\Pi}^3\tan^2\beta
\alpha_{s}^2}{4\pi^3 v_{w}^2}|C^*_0|^2,$$

$$\Gamma_{\Pi_t^0\rightarrow{\gamma Z}}=\frac{\alpha_e^2(1-\varepsilon)^2\tan^2\beta
M_{\Pi}^3}{9\pi^3 v_{w}^2}\tan^2\theta_w
(1-\frac{m_t^2}{M_{\Pi}^2})^2 L^2(M_\Pi),$$

with
$$C^*_{0}=C_{0}(-p_{\Pi},p_{g},m_{t},m_{t},m_{t}),$$
$$L(M_\Pi)=\int_0^1dx\int_0^1dy[1+(\frac{M_{\Pi}}{m_t})^2x(x-1)y^2+(\frac{M_Z}{m_t})^2
yx(y-1)]^{-1},$$ here,  $p_{\Pi}$ and $p_{g}$ denote the momenta
of $\Pi_t^0$ and one gluon, respectively.

After calculating the cross section
$\hat{\sigma}_{\gamma\gamma\rightarrow\Pi_t^0}(\hat{s})$ for the
subprocess $\gamma\gamma\rightarrow\Pi_t^0$, we can obtain the
total cross section at the $e^+e^-$ linear collider by folding
$\hat{\sigma}_{\gamma\gamma\rightarrow\Pi_t^0}(\hat{s})$ with the
photon distribution function $f_\gamma(x)$:
\begin{eqnarray}
\sigma_{total}(s)&=&\int_{\tau_{min}}^{\tau_{max}}
d\tau\int_{\tau/x_{max}}^{x_{max}}\frac{dx}{x}f_{\gamma}(x)
f_{\gamma}(\frac{\tau}{x})\hat{\sigma}_{\gamma\gamma\rightarrow\Pi_t^0}(\hat{s}).
\end{eqnarray}
Here, we denote $\tau=\hat{s}/s$ with $s$ being the $e^+e^-$ c.m.
energy squared. The photon distribution can be written as
\cite{distribution}:
\begin{equation}
f_{\gamma}(x)=\frac{1}{D(\xi)}[1-x+\frac{1}{1-x}-\frac{4x}{\xi(1-x)}+
\frac{4x^2}{\xi^2(1-x)^2}],
\end{equation}
 and
\begin{equation}
D(\xi)=(1-\frac{4}{\xi}-\frac{8}{\xi^2})\ln(1+\xi)+\frac{1}{2}+\frac{8}{\xi}-\frac{1}{2(1+\xi)^2}.
\end{equation}
Where, $\xi=\frac{4E_e\omega_0}{m_e^2}$. $E_e$, $\omega_0$ are the
incident electron energy and the laser-photon energy.
$x=\omega/E_e$ stands for the fraction of energy of the incident
electron carried by the back-scattered photon. $f_\gamma$ vanishes
for $x>x_{max}=\omega_{max}/E_e=\xi/(1+\xi)$. In order to avoid
the creation of $e^+e^-$ pairs by the interaction of the incident
and back-scattered photons, we require $\omega_0x_{max}\leq
m_e^2/E_e$ which implies that $\xi\leq 2+ 2\sqrt{2}\approx 4.8$.
Taking $\xi=4.8$,  we obtain $x_{max}\approx 0.83,D(\xi)\approx
1.8 $. Therefore, we can obtain the up-limit and down-limit of
integral as: $x_{max}=0.83$, $x_{min}=\tau/x_{max}$,
$\tau_{max}=x_{max}^2$, $\tau_{min}=M_{\Pi}^2/s$.

\section{The numerical results and discussion}

To obtain the numerical results, we take $m_t=178$ GeV, $v_t=60$
GeV. From the one-loop evaluation formula, we can obtain the value
of $\alpha_e$ at energy scale of ILC. Therefore, there leave three
free parameters: $\varepsilon$, $e^+e^-$ c.m. energy $\sqrt{s}$
and the mass of top-pion $M_{\Pi}$. To see the effect of
$M_{\Pi}$, $\varepsilon$ and $\sqrt{s}$ on the cross section, we
plot, in Fig.1-3, the cross section $\sigma_{total}$ as a function
of $M_{\Pi}$ with $\varepsilon=0.03,0.06,0.1$ and
$\sqrt{s}=500,800,1600$ GeV, respectively. We can see that the
cross section is at the level of tens fb to a hundred fb. There
exists a peak in the plot when $M_{\Pi}$ is near $2m_t$ which
arises from the top quark triangle loop. On the other hand, it is
shown that the increasing of $\sqrt{s}$ can depress the cross
section.

The neutral top-pion should be detected via its decay modes. We
know that the possible decay modes of $\Pi_t^0$ are $t\bar{t}$(if
$M_{\Pi}>2m_t$), ${t\bar{c}}$, $b\bar{b}$, $\gamma\gamma$, $gg$,
$\gamma{Z}$. The tree level decay modes $t\bar{t}$, $t\bar{c}$
$b\bar{b}$ and the loop level decay mode $gg$ are the main decay
modes of $\Pi_t^0$. The decaying branching ratios of the main
decay modes are shown in Fig.4. We can see that $\Pi_t^0$ almost
decays to $t\bar{t}$ when $M_{\Pi}>2m_t$. The decay branching
ratio of $\Pi^0_t\rightarrow t\bar{c}$ is the largest one when
$t\bar{t}$ mode is forbidden. We focus on studying the potential
to observe the $\Pi_t^0$ via these main decay modes. The event
number of the signal via each mode can be obtained via the formula
$N_X=L_{ee}\sigma_{total}Rb(\Pi_t^0\rightarrow X)$. The event
number plot versus $M_{\Pi}$ is shown in Fig.5 with $\sqrt{s}=800$
GeV,  $\varepsilon=0.06$ and the yearly luminosity $L_{ee}=500
fb^{-1}$. We can see that the event number of signal
$t\bar{c},b\bar{b},gg$ is significantly depressed when
$\Pi_t^0\rightarrow t\bar{t}$ is open. There are $10^3-10^4$
$t\bar{t}$ events can be produced for heavy
$\Pi_t^0(M_{\Pi}>2m_t)$. The total cross section of $\gamma\gamma
\rightarrow t\bar{t}$ is at the level of $10^2 fb$ at TeV energy
scale\cite{SM-section} in the SM. The number of $t\bar{t}$ event
produced via heavy $\Pi^0_t$ decaying is comparable to that in the
SM. Such $\Pi_t^0$ contribution to the $t\bar{t}$ production
should be easily detected at the planned ILC. Therefore, the
significant deviation of $t\bar{t}$ production from the SM
prediction might provide the clue of $\Pi_t^0$ existence. But it
is difficult to identify $\Pi_t^0$ via $t\bar{t}$ mode. The reason
is that the decay width of $\Pi_t^0\rightarrow t\bar{t}$ is very
large and the peak of $t\bar{t}$ invariant distribution is too
wide to be observed, we can not identify  $\Pi_t^0$ via the
$t\bar{t}$ invariant distribution. Therefore, we can conclude that
$t\bar{t}$ is an ideal mode to search for the clue of TC2 model
but it is not suitable to confirm the existence of heavy
$\Pi_t^0$. Because topcolor is non-universal, there exists the
flavor-changing coupling of the neutral top-pion to top and charm
quarks. So, $\Pi_t^0$ can decay to $t\bar{c}$ at tree-level which
is the typical feature of the TC2 model. Now, we discuss the
potential to observe $\Pi_t^0$ via $t\bar{c}$ mode. For the light
$\Pi_t^0$, the branching ratio of $\Pi_t^0\rightarrow t\bar{c}$ is
the largest one. Over $10^4$ $t\bar{c}$ events can be produced
with the integral luminosity $500 fb^{-1}$. For heavy $\Pi_t^0$,
the $t\bar{c}$ events significantly drop to $10^3$ with the
opening of $t\bar{t}$ mode. In order to obtain a sound conclusion
about the potential to observe the $\Pi_t^0$, we should consider
the SM background and b-tagging and c-tagging efficiency. If we
detect $\Pi_t^0$ via $t\bar{c}$, the only irreducible background
arises from $\gamma\gamma \rightarrow t\bar{c}$ in the SM, the SM
cross section of such process is much small(at the level of
$10^{-8} fb$\cite{rrtc}), therefore, the perfect identification of
$t\bar{c}$ can make the background become very clean. To identify
$t\bar{c}$, we should reconstruct top quark from its decay mode
$W^+b$. So, the b-tagging and c-tagging are need to identify
$t\bar{c}$. We take b-tagging efficiency as $60\%$ and c-tagging
efficiency as $35\%$\cite{efficiency}, there are also enough
$t\bar{c}$ events which can be identified. For example, we
typically take $t\bar{c}$ events as $10^4$, there are about
$2.1\times 10^3$ $t\bar{c}$ can be tagged, the corresponding
statistical uncertainty at the $95\%$ C.L. is $4.4\%$. So, with
the clean background, the large number of $t\bar{c}$ events tagged
and the observation of the peak of the $t\bar{c}$ invariant mass
distribution, we can obtain the clear signal of neutral top-pion
with high precision. As it is shown in Fig.5, a lot of  $gg$
events can also be produced via $\Pi_t^0\rightarrow gg$ which
leads to 2-jets. It is difficult to identify $\Pi_t^0$ via such
2-jets. So, $gg$ mode is not suitable to observe $\Pi_t^0$. For
the $b\bar{b}$ mode, there are over $10^2$ $b\bar{b}$ events can
be produced via light $\Pi_t^0$ decaying. The dominant background
to $\gamma\gamma\rightarrow \Pi_t^0 \rightarrow b\bar{b}$ comes
from the SM process $\gamma\gamma\rightarrow b\bar{b},c\bar{c}$.
Various techniques can be used to suppress these background. The
most effective technique appears to be to polarize the
initial-state photons: $\Pi_t^0$ are only produced from an initial
state with spin $J_z=0$, whereas, the leading-order
$b\bar{b}(c\bar{c})$ background are dominantly produced from the
$J_z=\pm 2$ initial state. More precisely, production from the
$J_z=0$ state is suppressed by $m^2_q/s$\cite{bb}. Hence,  in the
region of the $\Pi_t^0$ resonance, the $b\bar{b},c\bar{c}$
background are heavily suppressed if we choose initial state with
$J_z=0$. On the other hand, the processes $\gamma\gamma
\rightarrow b\bar{b}g(c\bar{c}g)$ can mimic a two-jet event. So,
such processes in the SM are also the important background. It is
shown that the $m^2_q/s$ suppressions do not necessarily apply to
the QCD corrected processes $\gamma\gamma \rightarrow
b\bar{b}g(c\bar{c}g)$ and the largest background is from the
$\gamma\gamma \rightarrow c\bar{c}g$. The existence of
$\gamma\gamma \rightarrow b\bar{b}g(c\bar{c}g)$ background make
the potential to observe $\Pi_t^0$ via $b\bar{b}$ mode become
unclear, and a detail study is needed which is beyond this paper.
Furthermore, in the SM or the MSSM, there exists the similar
process $\gamma\gamma \rightarrow H\rightarrow b\bar{b}$.
Therefore, to identify the $\Pi_t^0$ via $b\bar{b}$ mode, we
should also find the different feature between $\gamma\gamma
\rightarrow \Pi_t^0 \rightarrow b\bar{b}$ and $\gamma\gamma
\rightarrow H\rightarrow b\bar{b}$. From above discussion, we can
conclude that $\gamma\gamma \rightarrow \Pi_t^0 \rightarrow
t\bar{c}$ is an ideal mode to detect neutral top-pion which can
provide enough typical $\Pi_t^0$ signals.

In summary, the running of ILC with high energy and luminosity
will provide a good chance to test the new physics models. With
the realization of photon-photon collision, the neutral top-pion
can be produced as s-channel resonances in the process
$\gamma\gamma \rightarrow \Pi^0_t$ which is the most important
production mechanism of neutral top-pion. In this paper, we
calculate the cross section of the process $\gamma\gamma
\rightarrow \Pi^0_t$ and discuss the potential to observe the
neutral top-pion via its various decay modes. The result shows
that the cross section is at the level of tens fb to a hundred fb.
The possible decay modes of $\Pi_t^0$ are $t\bar{t}$(if
$M_{\Pi}>2m_t$), ${t\bar{c}}$, $b\bar{b}$, $\gamma\gamma$, $gg$,
$\gamma{Z}$. For the mode $\Pi_t^0\rightarrow t\bar{c}$, there are
$10^3-10^4$ $t\bar{c}$ events can be produced with the integral
luminosity $L_{ee}=500 fb^{-1}$ and the SM background is very
clean. So, $\Pi_t^0\rightarrow t\bar{c}$ is the most ideal mode to
search for neutral top-pion.

\section{Acknowledgments}

This work is supported by the National Natural Science Foundation
of China(Grant No.10375017), the Excellent Youth Foundation of
Henan Scientific Committee(Grant No. 02120000300), and the Henan
Innovation Project for University Prominent Research Talents(Grant
No. 2002KYCX009).

\newpage

\begin{figure}[htb]
\begin{center}
\epsfig{file=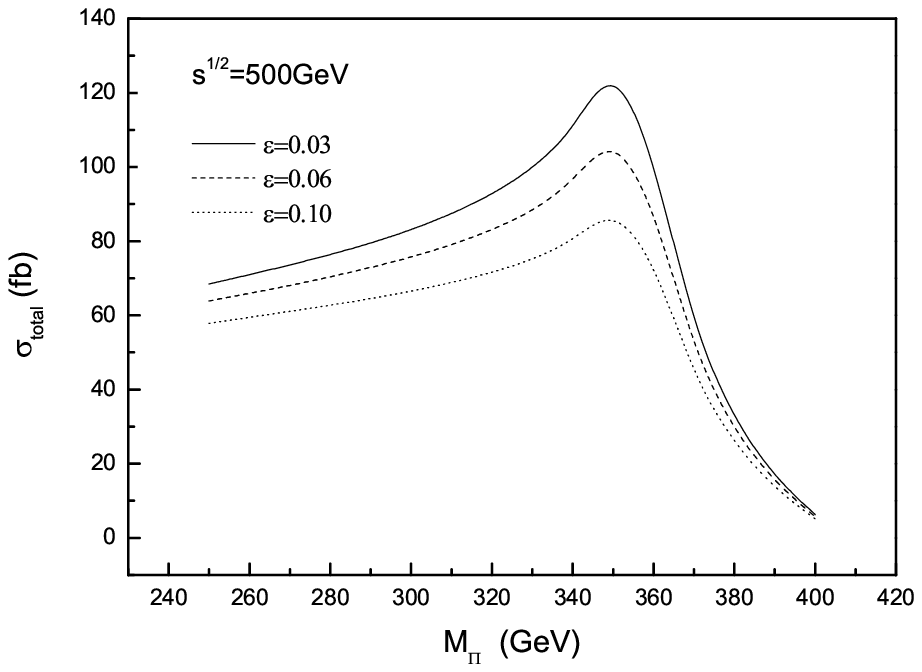, width=350pt,height=290pt}\vspace{1cm}
 \caption{The cross section of $\gamma\gamma\rightarrow\Pi_t^0$
 as functions of the top-pion mass $M_{\Pi}$ with
$\sqrt{s}=500$ GeV and $\varepsilon=$0.03, 0.06, 0.1.}
 \label{Fig.1}
\end{center}
\end{figure}

\newpage
\begin{figure}[htb]
\begin{center}
\epsfig{file=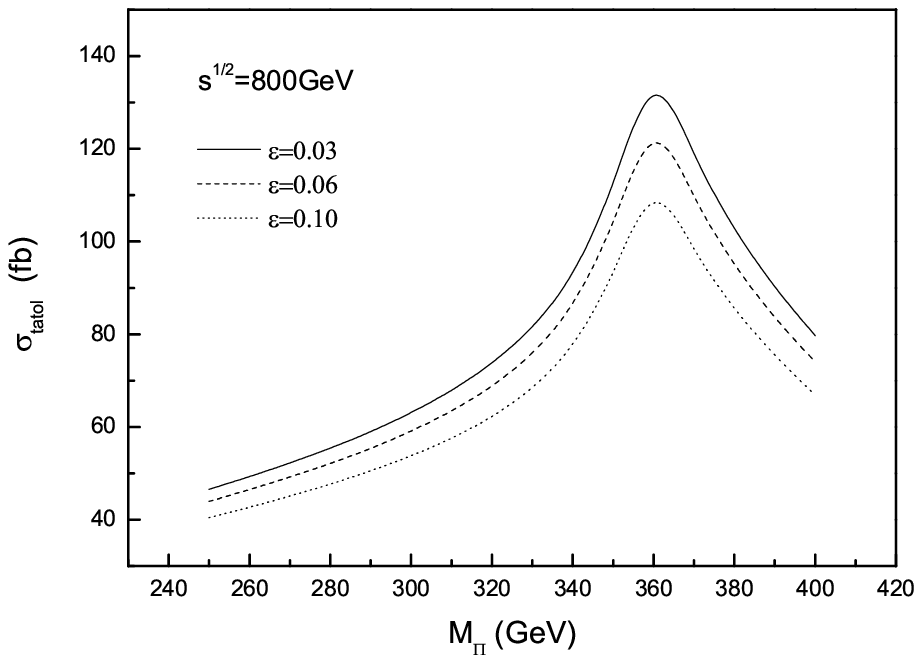, width=350pt,height=290pt}\vspace{1cm}
 \caption{The cross section of $\gamma\gamma\rightarrow\Pi_t^0$
 as functions of the top-pion mass $M_{\Pi}$ with
$\sqrt{s}=800$ GeV and $\varepsilon$=0.03, 0.06, 0.1.}
 \label{Fig.2}
\end{center}
\end{figure}

\newpage
\begin{figure}[htb]
\begin{center}
\epsfig{file=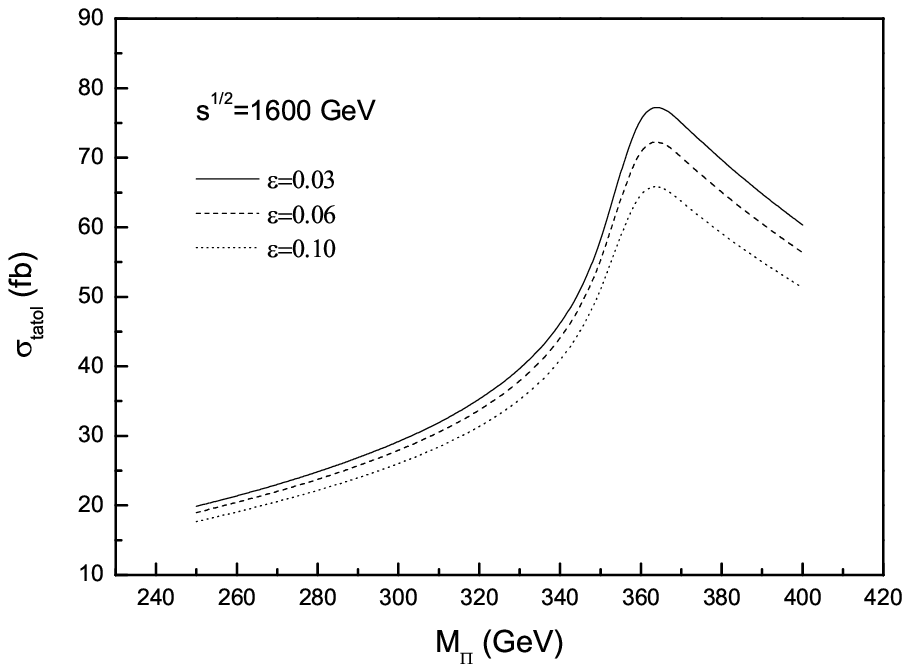, width=350pt,height=290pt}\vspace{1cm}
 \caption{The cross section of $\gamma\gamma\rightarrow\Pi_t^0$
 as functions of the top-pion mass $M_{\Pi}$ with
$\sqrt{s}=1600$ GeV and $\varepsilon$=0.03, 0.06, 0.1.}
 \label{Fig.3}
\end{center}
\end{figure}

\newpage
\begin{figure}[htb]
\begin{center}
\epsfig{file=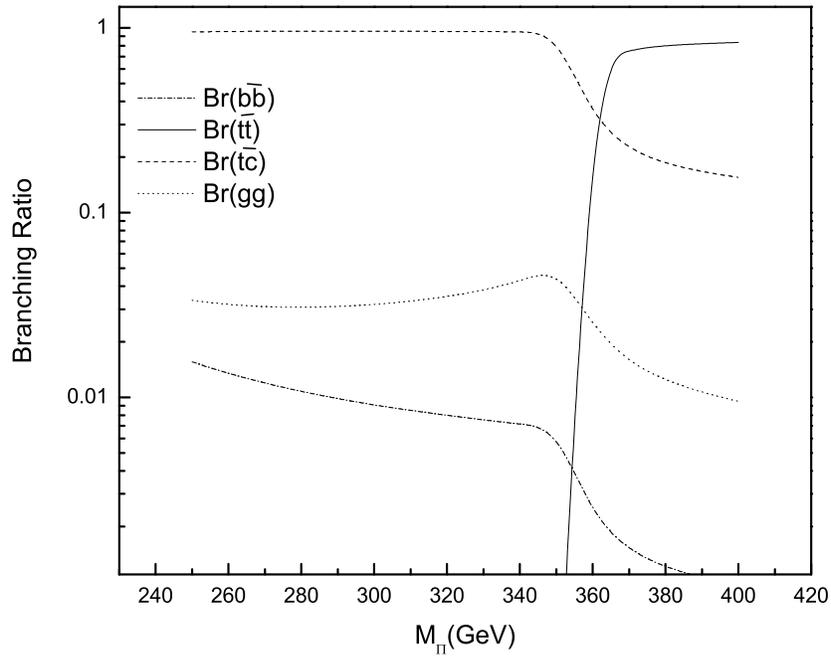, width=350pt,height=290pt}\vspace{1cm}
 \caption{The decaying branching ratios of $\Pi_t^0\rightarrow t\bar{t}, t\bar{c}, b\bar{b}, gg$
 as functions of the top-pion mass $M_{\Pi}$ with $\varepsilon=0.06$.}
 \label{Fig.4}
\end{center}
\end{figure}

\newpage
\begin{figure}[htb]
\begin{center}
\epsfig{file=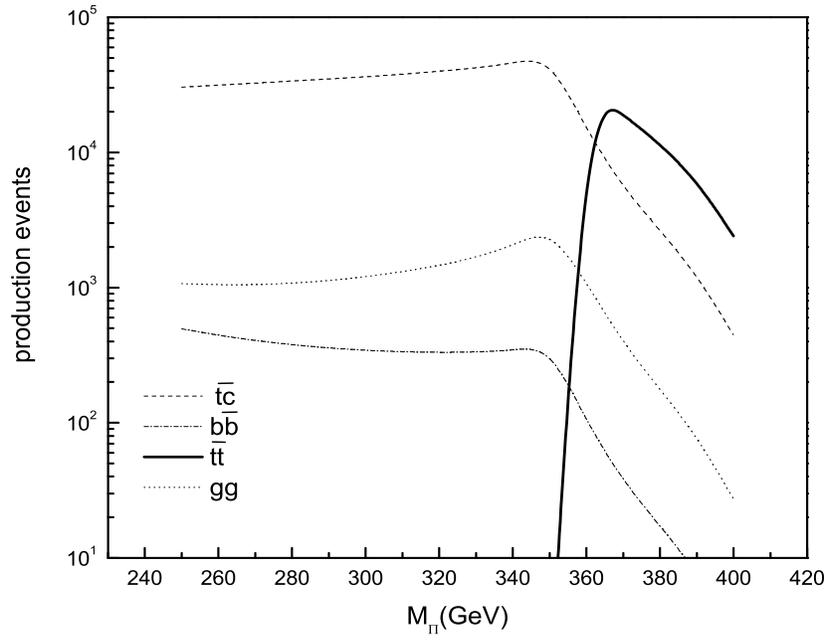, width=350pt,height=290pt}\vspace{1cm}
 \caption{The production event numbers of signal via the decay modes: $t\bar{t},
t\bar{c}, b\bar{b}, gg$ as
 functions of $M_{\Pi}$ with $\sqrt{s}=800$ GeV and $\varepsilon$=0.06.}
 \label{Fig.5}
\end{center}
\end{figure}

\end{document}